\documentclass[aps,twocolumn,prl,preprintnumbers,amsmath,amssymb,superscriptaddress,floats]{revtex4-1}

\usepackage{graphicx}% Include figure files
\usepackage[]{natbib}

\begin{document}

%\linenumbers
\title{Time-resolved terahertz dynamics in thin films of the topological insulator Bi$_{2}$Se$_3$}

\author{R. Vald\'es Aguilar}
\email{rvaldes@physics.osu.edu}
\affiliation{Center for Emergent Materials, Department of Physics, The Ohio State University, Columbus, OH 43210}
\affiliation{Center for Integrated Nanotechnologies, Los Alamos National Laboratory. MS K771. Los Alamos, NM 87545}
\author{J. Qi}
\affiliation{Center for Integrated Nanotechnologies, Los Alamos National Laboratory. MS K771. Los Alamos, NM 87545}
\author{M. Brahlek}
\affiliation{Department of Physics and Astronomy, Rutgers the State University of New Jersey. Piscataway, NJ 08854}
\author{N. Bansal}
\affiliation{Department of Physics and Astronomy, Rutgers the State University of New Jersey. Piscataway, NJ 08854}
\author{A. Azad}
\affiliation{Center for Integrated Nanotechnologies, Los Alamos National Laboratory. MS K771. Los Alamos, NM 87545}
\author{J. Bowlan}
\affiliation{Center for Integrated Nanotechnologies, Los Alamos National Laboratory. MS K771. Los Alamos, NM 87545}
\author{S. Oh}
\affiliation{Department of Physics and Astronomy, Rutgers the State University of New Jersey. Piscataway, NJ 08854}
 \author{A.J. Taylor}
\affiliation{Center for Integrated Nanotechnologies, Los Alamos National Laboratory. MS K771. Los Alamos, NM 87545}
\author{R.P. Prasankumar}
 \affiliation{Center for Integrated Nanotechnologies, Los Alamos National Laboratory. MS K771. Los Alamos, NM 87545}
\author{D.A. Yarotski}
\email{dzmitry@lanl.gov}
\affiliation{Center for Integrated Nanotechnologies, Los Alamos National Laboratory. MS K771. Los Alamos, NM 87545}
%\date{\today}

\begin{abstract}
We use optical pump--THz probe spectroscopy at low temperatures to study the hot carrier response in thin Bi$_2$Se$_3$ films of several thicknesses, allowing us to separate the bulk from the surface transient response. We find that for thinner films the photoexcitation changes the transport scattering rate and reduces the THz conductivity, which relaxes within 10 picoseconds (ps). For thicker films, the conductivity increases upon photoexcitation and scales with increasing both the film thickness and the optical fluence, with a decay time of approximately 5 ps as well as a much higher scattering rate. These different dynamics are attributed to the surface and bulk electrons, respectively, and demonstrate that long-lived mobile surface photo-carriers can be accessed independently below certain film thicknesses for possible optoelectronic applications.
\end{abstract}

\maketitle

Topological insulators (TI) represent a new state of matter \cite{RevModPhys.82.3045,RevModPhys.83.1057,annurev-conmatphys-062910-140432} and promise numerous applications in optoelectronics and spintronics.\cite{Pesin-review-spintronics,PhysRevLett.100.096407,PhysRevB.82.245107} In an ideal TI, insulating bulk does not contribute to the charge transport, and conductivity is 
determined solely by surface carriers. The surface state is topologically protected from backscattering and exhibits very low transport scattering rates (the inverse transport lifetime). For example, measurements of AC conductivity of thin films \cite{Rolando-Kerr, Rolando-NaturePhys} and single crystals\cite{Laforge,Sushkov} of exemplary TI material, Bi$_2$Se$_3$, found surface scattering rates of at most 2 THz which indicated, at least partially, the effect of the topological protection of the surface state.

However, the existing generation of TI materials, specifically Bi$_2$Se$_3$, exhibit significant contributions from the bulk carriers to the total conductivity. One reason is the chemical potential shift into the conduction band due to electron doping, typically by Se vacancies and anti-site defects.\cite{Se-vac} Another effect interfering with ideal TI response occurs when the bulk bands cross the Fermi level near the material surface (band bending) and generate two dimensional electron gas (2DEG). This 2DEG is confined within $\sim$20 nm from the surface \cite{Bianchi-NatCom,King-PRL} and coexists with the topological surface states. Although such 2DEG has been observed in single crystals,\cite{Bianchi-NatCom,King-PRL} there is no definite evidence for its existence in thin films of Bi$_2$Se$_3$. As we will show below, we observed significant changes in the transport dynamics even as we varied the film thickness well below the range expected for the effects of the 2DEG to occur, which suggests that it does not play a significant role in our measurements.

Optoelectronic functionality of TI relies on the dynamic response of non-equilibrium charge carriers in topologically protected surface states, making it important to unambiguously separate surface response from the interfering bulk. Although electronic measurements on FET structures have unveiled a plethora of TI transport parameters (scattering rate, plasma frequency, etc.), they cannot reliably distinguish signatures of the surface carriers from that of the bulk carriers present in non-ideal TI. Optical pump--optical probe spectroscopy (OPOP)~\cite{Jingbo-OPOP,Kumar-OPOP,Hsieh-PRL-2011} and time-resolved angle resolved photoemission spectroscopy (tr-ARPES)\cite{Sobota-trARPES,Wang-trARPES,Crepaldi-trARPES2013} were used to reveal dominant role of electron-phonon scattering in the carrier relaxation process following photoexcitation of single Bi$_2$Se$_3$ crystals. However, OPOP is most sensitive to interband processes, and tr-ARPES lacks high energy resolution near the chemical potential and is only sensitive to surface carriers. In contrast, time-averaging terahertz (1 THz$\sim$4 meV) time-domain spectroscopy (THz-TDS) and ultrafast optical pump-THz probe spectroscopy (OPTP) can directly interrogate the dynamics of the free carriers at the Fermi level that is relevant to transport properties, and can penetrate deep into the material, making it sensitive to both bulk and surface carriers. They are particularly useful for investigations of Dirac materials, as demonstrated in studies of the photoinduced carrier dynamics in graphene.\cite{Kampfrath-OPTP-g,George-OPTP-g,Choi-OPTP-g,Strait-OPTP-g,Jnawali-OPTP-g,Frenzel-OPTP-g} Recently, THz-TDS provided evidence of the surface carrier conductivity at THz frequencies with a relatively minor contribution from the bulk carriers in thin films of Bi$_2$Se$_3$.\cite{Rolando-NaturePhys,Rolando-Kerr} Here, we take advantage of the unique characteristics of OPTP spectroscopy to unravel the differences in the dynamics of hot surface and bulk carriers following photoexcitation in TI Bi$_2$Se$_3$. This will provide fundamental insight into the transport properties of photoexcited surface states, while laying the groundwork for their potential applications.

The thin film growth and characterization has been described elsewhere.\cite{film-growth} The three films, with thicknesses of 10, 16 and 20 QL (1 QL $\sim$ 0.94 nm), that we measured here were used before in THz-TDS experiments.\cite{Rolando-Kerr,Rolando-NaturePhys} These show an effective sheet carrier density of n$_{TI} \sim$1.5$\times$10$^{13}$ cm$^{-2}$ and a mobility of $\mu_{TI} \sim$ 900 cm$^2$/Vs.\cite{Rolando-Kerr} At THz frequencies the equilibrium response of these films is mostly thickness independent, with a scattering rate ranging between 1 and 1.5 THz (larger for thinner films).\cite{Rolando-Kerr} A contribution from bulk carriers could not be resolved, which meant that the scattering rate (mobility) was much larger (smaller) than in the surface states. Using the values of the THz plasma frequency and the linear dispersion of the topological surface states, we can estimate a value of the Fermi energy as E$_F \sim$ 0.45 eV.\cite{note1}

The OPTP experiments were performed using a home-built system. Pulses from an amplified Ti:sapphire laser system of 120 fs duration at 800 nm (1.55 eV) with a repetition rate of 1 kHz were split into 3 paths: 1) the pump to photoexcite the sample, 2) THz generation via optical rectification in a ZnTe crystal, and 3) a gate pulse for electro-optical THz detection in another ZnTe crystal. The pump beam was chopped at 500 Hz and the output of the balanced photodiode was fed to a transimpedance amplifier and then to a lock-in amplifier. The chopper was placed in either the gating beam path to measure the frequency-resolved THz spectra or in the pump beam to measure the frequency-integrated or frequency-resolved photoinduced signal at different pump-probe time delays $\tau_{pp}$. The pump fluence was varied between 3 and 200 $\mu$J/cm$^2$ using neutral density filters. Both the optical and THz beams were linearly polarized parallel to each other.

\begin{figure}
\includegraphics[width=8.5cm,keepaspectratio=true]{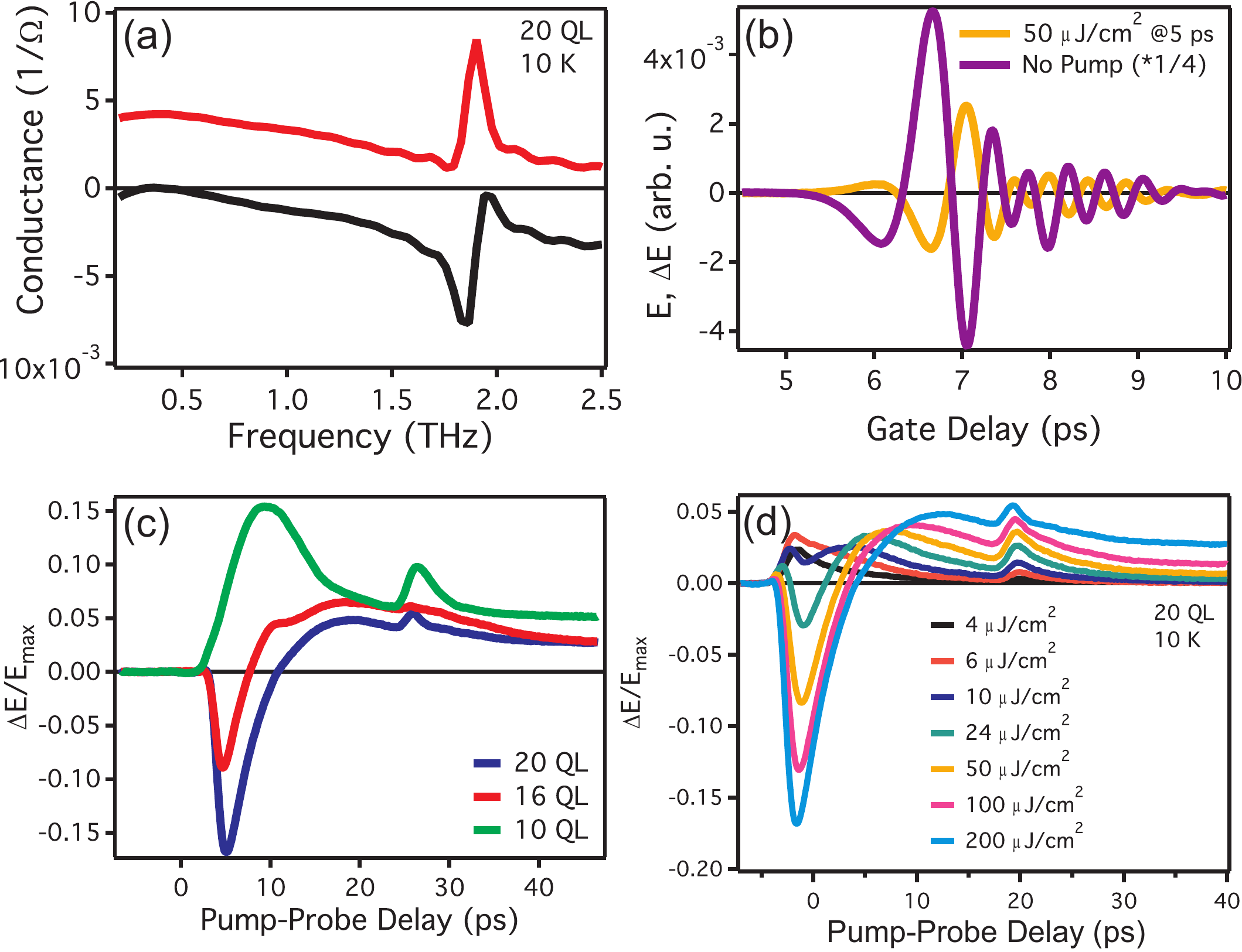}
\caption{Equilibrium and frequency-integrated transient THz responses of Bi$_2$Se$_3$ thin films at 10 K: (a) Equilibrium complex THz conductance of a 20 QL film. (b) Transmitted $E_T$ without the optical pump, and the photoinduced changes $\Delta E_T$ at $\tau_{pp}=5$ ps and a fluence of 50 $\mu$J/cm$^2$. The former trace is divided by 4 for ease of comparison. (c) Normalized $\Delta E_T$ as a function of $\tau_{pp}$ for 3 different samples (10, 16 and 20 QL) at a fixed pump fluence of 200 $\mu$J/cm$^2$. (d) Same as (c) for the 20 QL film at fluences between 4 and 200 $\mu$J/cm$^2$.}
\label{fig1}
\end{figure}

We have used this setup to measure the equilibrium complex conductance $G = \sigma_{xx} t$ (where $t$ is the film thickness) of all three samples. The results (dark line in Fig. \ref{fig1}b shows transmitted THz electric field $E_T$, and Fig. \ref{fig1}a shows real and imaginary parts of $G$) were essentially identical to previous reports \cite{Rolando-Kerr,Rolando-NaturePhys} even though the films have been kept under atmospheric conditions for months (stored inside a dry-box with relative humidity lower than 20 \%). This shows the robustness of the THz response expected for a topologically protected surface state. The only significant difference is a larger value of the scattering rate, which implies larger surface disorder.

When the sample is excited by the pump pulse, there is a change in the transmitted THz electric field, shown as $\Delta E_T$ in Fig. \ref{fig1}b (light line) at a fixed $\tau_{pp} = 5$ ps. It is important to highlight that the peak value of $\Delta E_T$ is displaced in time from the peak of the equilibrium $E_T$, signifying a change in the phase of the complex transmission function. This implies that both the real (modifies the amplitude) and imaginary (modifies the phase) parts of the $G$ are changing after photoexcitation.

We clarify this behavior by measuring frequency-integrated $\Delta E_T$ as a function of $\tau_{pp}$ by keeping the gate delay at the peak of equilibrium THz field. For a pump fluence of 200 $\mu$J/cm$^2$ as shown in Fig. \ref{fig1}c, the thinner film shows only a positive $\Delta E_T$, i.e. the film becomes more transparent, whereas the thicker films have an initial rapid drop in transmission and then become more transparent, with a relaxation time similar to the thinner film. Around $\tau_{pp}\sim$25 ps there is an additional peak present in $\Delta E_T$ for all films, which comes from a back--reflection of the pump beam in the substrate. Although the thinnest film becomes more transparent for all the pump fluences used here, there is an interesting fluence dependence of the $\Delta E_T$ in the thicker films. Figure \ref{fig1}d shows the full fluence dependence of the 20 QL film (similar behavior is observed for the 16 QL film). At low fluences the response is essentially the same as the thinner film, but as the fluence increases the film becomes more opaque for a short period of time and then the response returns to that of the thinner film. Photoexcitation at high pump fluences thus appears to create carriers of a fundamentally different nature in thick and thin films, even at the same fluences. The response of the thicker films depends on their thickness, which suggests that the photoexcited carriers originate primarily from the bulk.

\begin{figure}[t]
\centering
\includegraphics[width=6cm,keepaspectratio=true]{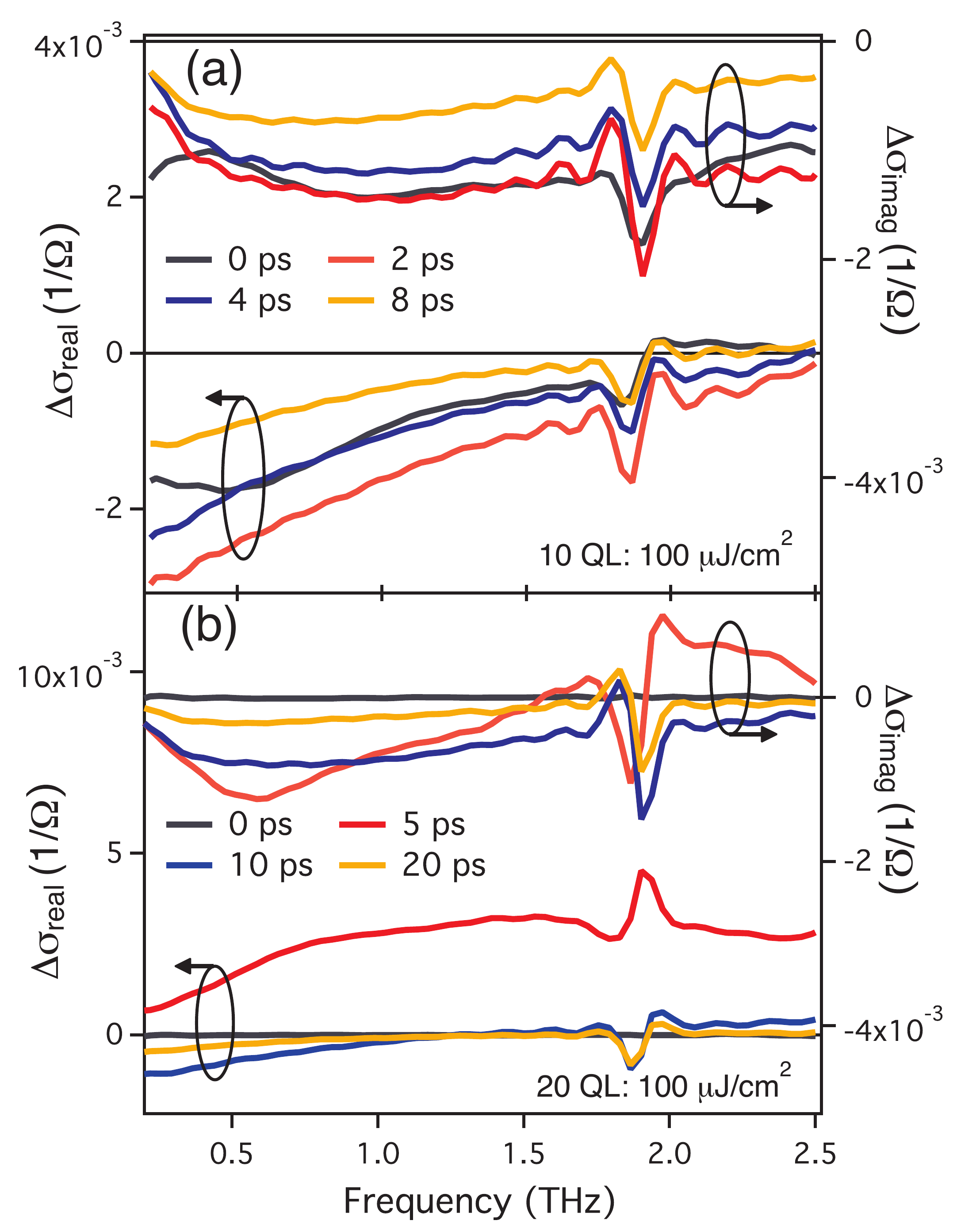}
\caption{Photoinduced changes in the complex conductivity of the (a) 10 QL and (b) 20 QL films at 10 K for several $\tau_{pp}$ at a pump fluence of 100 $\mu$J/cm$^2$. Circles and arrows indicate the appropriate axes to use for real and imaginary parts of conductance.}
\label{fig2}
\end{figure}

\begin{figure}[t]
\centering
\includegraphics[width =6cm,keepaspectratio=true]{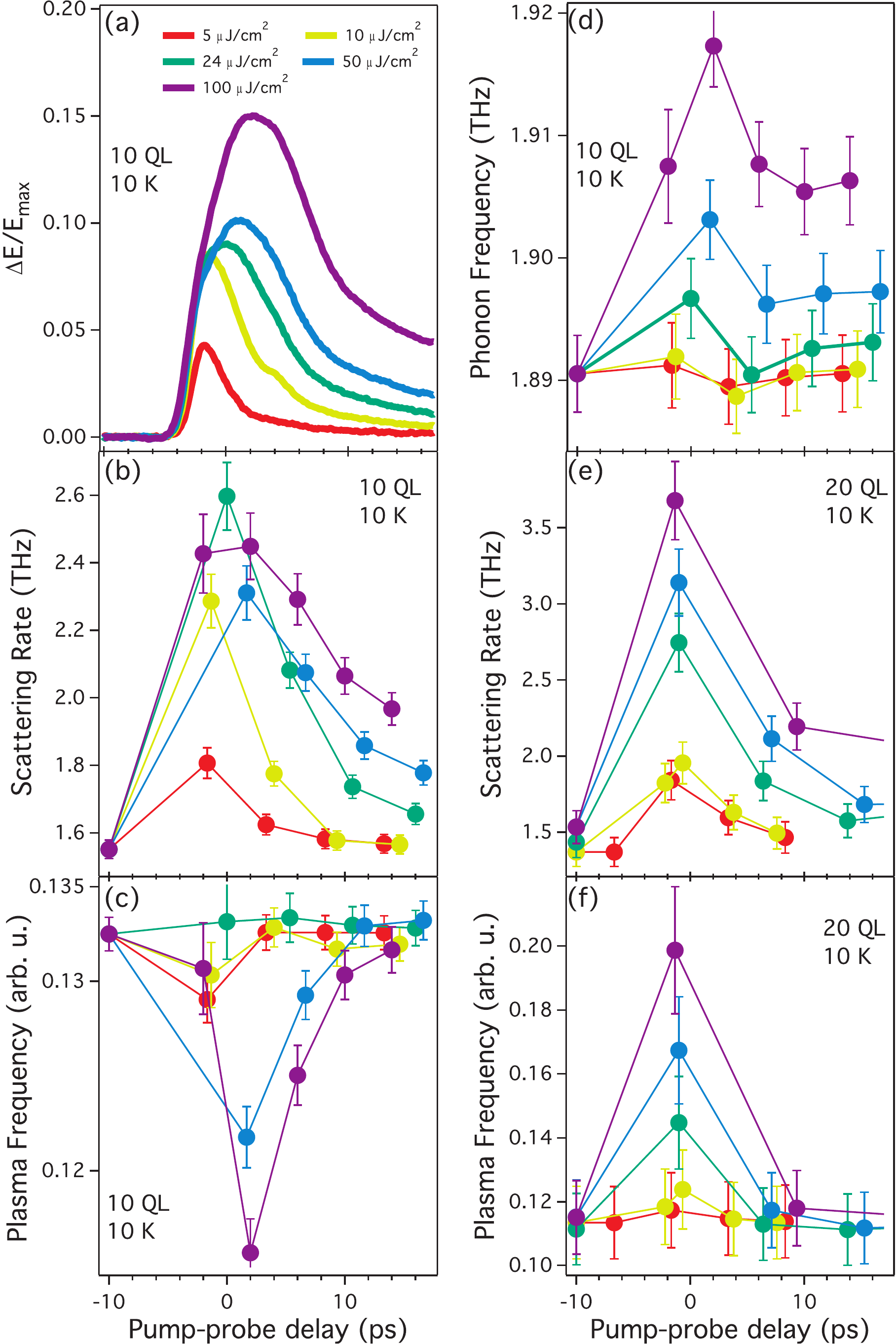}
\caption{Dynamics of Drude and phonon fit parameters: (a) Frequency-integrated $\Delta E_T$ of the 10 QL film at several pump fluences. (b) and (c) The time dependence of the $\Gamma_D$ and $\omega_{p}$, respectively, of the 10 QL film at the same fluences. (d) $\omega_o$ of the bulk optical phonon in the 10 QL film. Note the resemblance of the fluence dependence of the $\omega_o$ in (d) and $\Delta E_T$ in (a). $\Gamma_D$ (e) and $\omega_{p}$ (f) of the 20 QL film. In all panels, the error bars signify the parameter interval where the fit does not change.}
\label{fig3}
\end{figure}

To understand this behavior we obtained the frequency-resolved changes in conductivity $\Delta\sigma$ for several $\tau_{pp}$, starting with the 10 QL sample shown in Fig. \ref{fig2}a. From the negative values of $\Delta\sigma_{real}$ (left axis) we can see that the absorption decreases at low frequencies. This explains the initial increase in transparency revealed in the frequency-integrated response shown in Fig. \ref{fig1}c. We note that $\Delta\sigma_{imag}$ is also negative (right axis in Fig. \ref{fig2}a) further confirming that the film becomes more insulating.
 Another aspect to note in the spectra is a differential peak located near the optical phonon resonance slightly below 2 THz, which implies a slight shift of the resonance to higher frequencies (Fig. \ref{fig3}d).

The response of 20 QL film at large $\tau_{pp}$ is qualitatively similar to that of the 10 QL film (Fig. \ref{fig2}b), i.e. a negative change in both real and imaginary parts of the conductivity. However, at early times ($\tau_{pp}\sim$5 ps) there is a large positive change, i.e. an increase of the $\Delta\sigma_{real}$. This causes the fast negative change in the frequency-integrated $\Delta E_T$ in Fig. \ref{fig1}c and d. We note that this increase in conductivity has a large frequency extent, going beyond the experimental high frequency detection limit. This again indicates that the positive contribution must have a different origin than the equilibrium response or the pump-induced change in conductivity of the thinner film. Finally, we also detect a time-dependent change in the phonon frequency, very similar to that in the 10 QL film.

We performed fits of the $\Delta\sigma$ spectra using the same parametrization of the THz response as was used in previous studies.\cite{Rolando-Kerr,Rolando-NaturePhys} In equilibrium, this parametrization includes: 1) a single Drude free electron response whose spectral weight is specified by a plasma frequency $\omega_{p}$, and a scattering rate $\Gamma_D$ (inverse of the transport lifetime); 2) a Drude-Lorentz oscillator that reproduces the phonon resonance with its own plasma frequency $\omega_{pp}$, inverse lifetime $\Gamma_{p}$ and center frequency $\omega_o$; and 3) an effective background dielectric constant $\varepsilon_\infty$ that characterizes the effect of the high energy optical transitions on the low frequency electrodynamics and only appears in the imaginary part of the conductivity as a linear frequency-dependent component with a negative slope. An attempt to fit the Drude response with two different sets of carriers (bulk and surface) has not resulted in a unique set of parameters, hence a single Drude term is used in the presented fit. We start with the data from the 10 QL film shown in Figs. \ref{fig3}a,b and c. In this case $\Gamma_D$ shows a similar time dependence as the $\Delta E_T$. However, the $\omega_{p}$ does not change significantly even for the largest fluences. Based on the equilibrium carrier density and the largest change observed in the $\omega_{p}$, we estimate a maximum photoinduced change in the carrier density of 10\%, assuming that the effective mass of the carriers stays the same. Despite this increase in carrier density, the 10 QL film becomes more transparent, indicating reduced conductivity. This clearly demonstrates that an increase in $\Gamma_D$ dominates the photoinduced conductivity changes, as previously reported in graphene.\cite{Jnawali-OPTP-g,Frenzel-OPTP-g}

We now turn to the signature of the bulk phonon in the 10 QL film in Fig. \ref{fig3}d, where the phonon frequency $\omega_o$ versus $\tau_{pp}$ at all measured fluences is shown. We can see that $\omega_o$ increases after photoexcitation, resulting in the differential line shape shown in Fig. \ref{fig2}a. This increase in $\omega_o$ resembles the one observed in the temperature dependence of the equilibrium THz spectra,\cite{Laforge,Rolando-Kerr,Rolando-NaturePhys} which suggests that the lattice temperature rises and causes the increase in $\Gamma_D$, as was reported in the supplement to ref. \citet{Rolando-NaturePhys}. For the 20 QL film, the Drude parameters are shown in Fig. \ref{fig3}e and f. The $\Gamma_D$ has a time dependence similar to the thinner film. The most significant difference is in the behavior of $\omega_{p}$, which shows a very large enhancement that only survives for a few picoseconds, effectively duplicating the time dependence of the negative contribution to the $\Delta E_T$ in Fig. \ref{fig1}c. The largest photoexcited carrier density of $5\times10^{18}$/cm$^3$ (or a change of $\sim70$\%) is obtained at a fluence of 100 $\mu$J/cm$^2$. The 20 QL film thus becomes more opaque because of the extra photoexcited carriers that absorb and screen the THz probe pulse. We note here that the value of $\Gamma_D$ of the photoexcited carriers in the 20 QL film is similar ($>$2.5 THz at high fluences) to that obtained after the transition to a trivial (non-TI) state in similar films.\cite{Rolando-NaturePhys} This indicates the different nature of the photoexcited carriers from the equilibrium surface state.

The observed thickness dependence of the transient THz response has to be reconciled with the thickness independence of the equilibrium THz conductivity. Thus, we consider the following scenario. At 800 nm the optical penetration depth is approximately 50 nm,\cite{Sobota-trARPES} which implies that the pump fluence varies only slowly over the thickness of our films, and therefore can be considered effectively uniform. In the low fluence regime, the photoinduced dynamics are thickness independent (see Figs \ref{fig1}d and \ref{fig3}a) as all films become more transparent due to increased $\Gamma_D$. This behavior is similar to the relaxation of the electronic temperature of the surface states found in tr-ARPES experiments (compare Fig. \ref{fig3}a here with Fig. 4a in~\citet{Wang-trARPES}). This implies that the photoexcited electrons thermalize to a high temperature distribution at timescales shorter than the OPTP experimental resolution ($\sim$1 ps), and the excess thermal energy is subsequently relaxed, judging by the timescales of the decay, via the scattering between electrons and acoustic phonons at the sample surface.\cite{Wang-trARPES} At higher fluences in the thinner film, thermal diffusion starts to dominate, as the thermal energy can only be carried away at the diffusion rate $D$ ($D \sim$ 1.2 cm$^2$/s~\cite{Kumar-OPOP}), which for a pump photon energy of 1.55 eV implies a timescale of $\sim$21 ps.\cite{Sobota-trARPES} This explains the large residual values of the $\Gamma_D$ and phonon frequency changes at longer delays. In the thicker films, thermal diffusion seems to also dominate at longer delays, but at shorter times there is a large contribution characterized by an increased plasma frequency that rises and decays at much shorter timescales than $\Gamma_D$. This is consistent with the bulk relaxation times ($\sim$2 ps) obtained in previous experiments.\cite{Jingbo-OPOP,Hsieh-PRL-2011,Sobota-trARPES,Wang-trARPES,Crepaldi-trARPES2013} Thus, the simplest way to interpret the observed behavior is that photoinduced change $\Delta\sigma$ of 10 QL film is produced by hot surface carriers, while in thicker films the majority of photoexcited carriers live in the bulk of the material, and survive for only a few picoseconds after which surface carrier start to dominate the THz transport.

In conclusion, we have shown that there are two types of photoexcited carriers contributing to conductivity of Bi$_2$Se$_3$ films. At low temperatures and low pump fluences for all film thicknesses, we only observed a short-living ($\sim$10 ps) increase in the transport scattering rate without a significant change in the carrier concentration. This continues to be true as the fluence is increased in the thinnest (10 QL) film hinting at the surface nature of the involved carriers. However, in thicker films there is an additional contribution with a faster rise and decay times ($\sim$5 ps). This contribution increases both with fluence and thickness and therefore must be associated with photoexcited bulk carriers. In addition, the photoinduced scattering rate is much higher at high fluences than at low fluences or that of equilibrium carriers, consistent with the shorter relaxation time of bulk carriers. These results agree well with recent observations by Sim et al. that bulk and surface hot carriers co-exist for a long time following photoexcitation at low temperatures unlike in high temperature regime where bulk carriers scatter to surface states within 4 ps.\cite{Choi-OPTP} However, our work go beyond their conclusions and show that it is possible to distinguish surface and bulk photoelectron dynamics using optical pump--THz probe spectroscopy with varying pump fluence even when bulk carriers are present in equilibrium, thus presenting intriguing possibilities for potential optoelectronic applications.\cite{Gabor04112011}

We thank P. Armitage, J. Furdyna and X. Liu for sharing samples. This work was supported by the Department of Energy, Office of Basic Energy Sciences, Division of Material Sciences and by the University of California, Office of the President's joint UC/Laboratory program.

\bibliography{OPTP_Bi2Se3_References}

\begin{thebibliography}{29}
\bibitem [{not()}]{note1}%
  \BibitemOpen
  \href@noop {} {}\bibinfo {note} {Although it is known that the dispersion of
  the surface states in Bi$_2$Se$_3$ is not linear at high energies,
  corrections from the linear dispersion only modify the value of the Fermi
  level by $\approx$ 10\%.}\BibitemShut {Stop}%
\end{thebibliography}
\end{document}